\documentclass[aps,pre,a4paper,twocolumn,groupedaddress,showpacs,showkeys,preprintnumbers,floatfix,dvips]{revtex4}

\usepackage[utf8]{inputenc}
\usepackage[T1]{fontenc}
\usepackage{times}
\usepackage{graphics}
\usepackage{psfrag}
\usepackage{color}
\usepackage{amsmath}
\usepackage{amscd}
\usepackage{amssymb}
\usepackage{subfig}
\usepackage[normalem]{ulem}
\usepackage{caption}
\usepackage{ragged2e}

\DeclareCaptionJustification{reallyjustified}{\justifying}
\begin{document}

\preprint{cond-mat/0602244}

\title{The Network of Epicenters of the Olami-Feder-Christensen Model of
  Earthquakes}


\author{Tiago P. Peixoto}
\email[]{tpeixoto@if.usp.br}
\author{Carmen P. C. Prado}
\email[]{prado@if.usp.br}

\affiliation{Instituto de F\'{i}sica, Universidade de S\~{a}o Paulo \\
  Caixa Postal 66318, 05315-970 - S\~{a}o Paulo - S\~{a}o Paulo - Brazil}


\date{\today}

\begin{abstract}
  We study the dynamics of the Olami-Feder-Christensen (OFC) model of
  earthquakes, focusing on the behavior of sequences of epicenters
  regarded as a growing complex network.  Besides making a detailed
  and quantitative study of the effects of the borders (the occurrence
  of epicenters is dominated by a strong border effect which does not
  scale with system size), we examine the degree distribution and the
  degree correlation of the graph. We detect sharp differences between
  the conservative and nonconservative regimes of the model. Removing
  border effects, the conservative regime exhibits a Poisson-like
  degree statistics and is uncorrelated, while the nonconservative has
  a broad power-law-like distribution of degrees (if the smallest
  events are ignored), which reproduces the observed behavior of real
  earthquakes. In this regime the graph has also a unusually strong
  degree correlation among the vertices with higher degree, which is
  the result of the existence of temporary attractors for the
  dynamics: as the system evolves, the epicenters concentrate
  increasingly on fewer sites, exhibiting strong synchronization, but
  eventually spread again over the lattice after a series of
  sufficiently large earthquakes.  We propose an analytical
  description of the dynamics of this growing network, considering a
  Markov process network with hidden variables, which is able to
  account for the mentioned properties.
\end{abstract}

\maketitle

\section{Introduction}

Several different phenomena in nature spontaneously exhibit scale
invariant statistics. An attempt to identify a supposed basic
mechanism behind this behavior was made by Bak et al~\cite{bak:1987},
who introduced the concept of Self-Organized Criticality (SOC). SOC is
characterized by slowly driven systems, with fast avalanche-like
bursts of dissipation. Despite probably not being the sole explanation
for scale-invariance in nature, a wide range of systems do appear to
exhibit SOC, such as sand piles~\cite{bak:1987}, forest
fires~\cite{drossel:1992} and earthquakes~\cite{olami:1992}. However,
no general framework for SOC systems exist, and the mechanism behind
it is nor very well understood. In particular, the existence of SOC in
nonconservative systems is still
debated~\cite{lise:2001,lise2:2001,carvalho:2000}. This discussion is
frequently focused on one of the most studied and archetypal
nonconservative SOC models, the Olami-Feder-Christensen (OFC) model
for earthquakes.  Despite being defined by very simple rules (see
section~\ref{sec:ofc}), this model possesses very rich dynamics, and
is able to reproduce a wide range of statistics of real earthquakes,
such as the Gutenberg-Richter law for the distribution of event
sizes~\cite{gutenberg:1956,olami:1992} and the Omori law for fore- and
aftershocks~\cite{omori:1894,hergarten:2002}.

In this work we concentrate on the behavior of the epicenters in the
OFC model, both in the conservative and nonconservative regime,
studied as a growing complex network with scale free
behavior~\cite{peixoto:2004, peixoto2:2004}.

As known previously~\cite{middleton:1995,lise:2001}, we confirm that
in both regimes epicenters are more frequent closer to the border,
and study this effect in detail.  We show, however, that this border
effect does not scale with system size, and should not therefore be
considered representative of the dynamics of the model in the
thermodynamic limit. The length of the effect is dependent on the
level of dissipation, and is relatively large for the range of
parameters normally studied, specially when close to the conservative
limit, where a exponentially-decaying layer dominates, and it is hard
to observe anything else other than this border effect.  The existence
of this non-scaling border is in accordance with what was found
in~\cite{lise:2001}, that only earthquakes from a smaller internal
subset of the lattice exhibits Finite-Size Scaling in the event size
statistics. 

We turn then to the dynamics of epicenters. Recently there has been an
increasing interest in Complex Networks~\cite{newman:2002} as a tool
for describing very diverse systems, many of which exhibit a type of
scale invariance, that seems to be due to a general mechanism of
preferential attachment~\cite{price:1976,barabasi:1999}. In order to
study the epicenter dynamics in the OFC model, we construct a network
of consecutive epicenters in the bulk, and examine its properties in
more detail~\footnote{In a previous study~\cite{peixoto:2004}, we also
  analyzed some aspects of the same network, but did not take into
  account the border effect, and looked only at smaller lattices.}.

The network of epicenters, in the nonconservative regime, shows scale
invariance in the degree statistics, if the epicenters of the smaller
events are discarded. This network has also an unusual correlation
among vertices of high degree, which makes it very distinct from
networks created with a preferential attachment rule. These results
reproduce what has been found by Abe and
Suzuki~\cite{abe:2004,abe:2006} for real earthquakes, further
contributing to the success of this simple model in capturing the
essential earthquake dynamics. We show that this degree correlation
seems to be due to the existence of temporary attractors for the
dynamics, that shows periods of strong synchronization. We also
noticed that a drop in the average in-degree of the network seems to
precede big earthquakes, what could in principle be used to predict at
least an increase in the probability of big events in a given fault. 

We also show that is possible to reproduce some of the characteristics
of the complex epicenter network found in the nonconservative version
of the OFC model defining a growing procedure based on a Markov Chain
with hidden variables. To each possible epicenter (vertex) is attached
a hidden variable, and the probability of connections among epicenters
(related to the time sequence of events) is now given as a function of
the hidden variable of both vertices (instead of a simple preferential
attachment rule, as in a Barab\'{a}si-Albert type network~\cite{barabasi:1999}). 

This paper is organized as follows: in section~\ref{sec:ofc} we
briefly present the Olami-Feder-Christensen model for earthquakes; in
section~\ref{sec:border} we discuss in detail the way the spatial
distribution of epicenters depends on the distance to the borders, in
the conservative, nonconservative and ``almost conservative'' regimes;
in section~\ref{sec:epicenter-graph} we review the way we can built a
scale free network from the time series of epicenters, and present the
main properties of this network when the border effect is discarded. 
This network, although showing a scale-free behavior, is quite
different from Barab\'{a}si-Albert type networks, with a strong
correlation among vertices with high degree.  In
section~\ref{sec:markov-graph} we show how we can grow a network with
similar properties based on a Markov Chain process with hidden
variables and finally, in section~\ref{sec:conclusions}, we summarize our results.

\section{The OFC model}
\label{sec:ofc}
The OFC model~\cite{olami:1992} was inspired by the Burridge-Knopoff
spring-block model~\cite{burridge:1967}, and is defined as a 2D coupled map on a
square lattice. To each site $(i,j)$ in the lattice is assigned a
``tension'' $z_{ij}$, initially chosen at random from the interval
$[0,z_c[$. The entire system is driven slowly, with every $z_{ij}$
increasing uniformly. Whenever a site reaches the threshold tension 
($z_{ij} = z_c$), an avalanche starts (the ``earthquake''). The first
site to reach $z_c$ and start an avalanche is called the epicenter. A
fraction $\alpha$ of the tension of the toppling site is transfered to
each of its four neighbors ($z_{i\pm 1,j\pm 1} = z_{i\pm 1,j\pm 1} +
\alpha z_{ij}$), and its tension is set to zero ($z_{ij}=0$). If any of
the neighbors acquires a tension $z_{i\pm 1,j\pm 1} \ge z_c$, the same
toppling rules are applied, until there are no more sites in the
system with $z_{ij} \ge z_c$. Without loss of generality, we set
$z_c=1$. The total number of sites that topple until the avalanche is
over is called the ``size'' of the avalanche. The parameter $\alpha$
defines the level of local conservation of the system. For
$\alpha=0.25$ the system is locally conservative and for $\alpha<0.25$
it is dissipative. We consider here only the case with open boundary
conditions, i.e., the sites at the border of the lattice transfer
tension to nonexisting neighbors, so the system is always globally
nonconservative, but tends to conservative in the
thermodynamic limit if $\alpha=0.25$.

\section{Influence of the borders in the frequency of epicenters}
\label{sec:border}

\begin{figure}
  \centering \psfrag{L=800}[cb][l][1.2]{$L=800$}
  \psfrag{L=400}[c][l][1.2]{$L=400$} 
  \psfrag{800}[r][r][1.2]{$L=800$}
  \psfrag{400}[r][r][1.2]{$L=400$} 
  \psfrag{s>1}[r][r][1.2]{$s \ge 2$}
  \psfrag{s>29}[r][r][1.2]{$s \ge 30$} 
  \psfrag{0.25}[c][c][1.5]{(a) $\alpha=0.25$} 
  \psfrag{0.249}[c][c][1.5]{(b) $\alpha=0.249$, $s\ge 2$} 
  \psfrag{0.18 2-inf}[c][c][1.5]{(c) $\alpha=0.18$, $s\ge 2$}
  \psfrag{0.18 30-inf}[c][c][1.5]{(d) $\alpha=0.18$, $s\ge 30$}
  \psfrag{0.22 2-inf}[c][c][1.5]{(e) $\alpha=0.22$, $s\ge 2$}
  \psfrag{0.22 30-inf}[c][c][1.5]{(f) $\alpha=0.22$, $s\ge 30$}
  \resizebox{1.0\columnwidth}{!}{\includegraphics{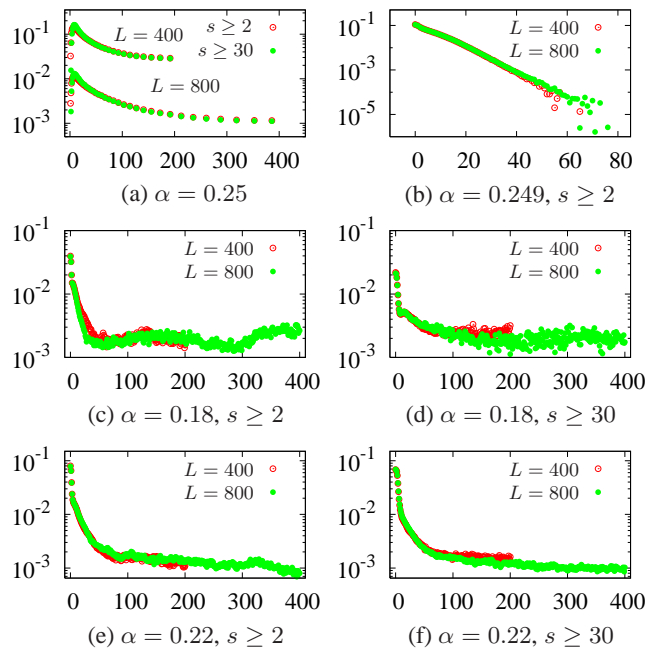}}
  \caption{ (color online) Frequency of epicenters (y axis) as a function of
    the distance from the border (x axis), for different values of $\alpha$,
    $L$ and earthquakes sizes $s$.  The data for $L=400$ in (a) was
    shifted upwards for clarity. From (c) to (f) the data for
    different values of $L$ were collapsed on top of each other by
    hand. All quantities are dimensionless.
    \label{fig:border}}
\end{figure}

We find that, in the stationary regime of the OFC model, the number of
times a site is an epicenter varies according to how close that site
is from the border, with epicenters closer to the border occurring
much more often. We will refer to this excess of epicenters in the
borders as the \emph{border effect}.  Fig.~\ref{fig:border} shows the
average frequency in which a site was an epicenter, given its distance
from the border, for $\alpha=0.25$, $\alpha=0.249$, $\alpha=0.22$ and
$\alpha=0.18$.  We have gathered statistics from two lattice sizes,
$L=400$ and $L=800$, and considered at least $6\times10^6$ events
(after the transient). We have considered epicenters only from
earthquakes larger than one ($s\ge2$), since size one earthquakes seem
to obey their own statistics~\cite{grassberger:1994}.  We have also
considered epicenters that gave rise to larger earthquakes ($s\ge30$),
to observe the dependence of the border effect to earthquake
size~\footnote{We realize however that this is rather coarse, since,
  due to the power-law distribution of event sizes, there is no
  characteristic event size to compare to. We wanted only to detect
  eventual differences in the statistics from the ``very small''
  events. It would also take a much longer time to consider only
  larger earthquakes.}.

For the conservative regime, as can be seen in
Fig.~\ref{fig:border}(a), the border effect is clearly weaker than
in the nonconservative regime, and the decay proceeds slowly towards
the bulk. It does not seem to scale with system size. Moreover, the
dependence on earthquake size appears to be weak for most of the
border effect, except for the very first few layers of sites close to
the border. 

Figures~\ref{fig:border}(c) to~\ref{fig:border}(f) also shows the same
results for the nonconservative case, for $\alpha=0.18$ and
$\alpha=0.22$. We notice that the border effect is composed roughly of
three parts: A thin region, comprised of the first few sites closest
to the border, where the effect is strong and seems to decay
exponentially.  This region is followed by a thicker layer of sites,
with a slower but also exponentially-decaying effect, and finally there is
a third region in which the decay is not exponential and proceeds
still more slowly towards the bulk of the system. None of this regions
seems to scale with system size, with the possible exception of the
third longer layer.  The overall border effect seems, however, to
depend on the earthquake size (on the contrary of what was observed in
the conservative case), as can be seen in Figs.~\ref{fig:border}(d)
and~\ref{fig:border}(f), which
shows clearly that the border effect decays more slowly towards the
bulk of the system if only larger events are considered. In Figs.~\ref{fig:border}(c)
to~\ref{fig:border}(f), the data for lattices of different
size $L$ was collapsed by hand, that is, curves were shifted up and
down in order to coincide, since statistics are different in each
case.  The slope and the size of the layers, however, were not
changed. 

The border effect also depends on $\alpha$. The closer the system is
to the conservative regime, the stronger and thicker is the layer of
sites affected by it. Note that for $\alpha=0.249$, the ``almost
conservative'' case (see Fig.~\ref{fig:border}(b)), the border effect
is so strong that almost no epicenters happen in the bulk of the
system, and only the fast exponentially-decaying border effect is
seen. This indicates that the lattice size $L=800$ is still too small
to study the system in this regime. If we compare this figure with
Fig.~\ref{fig:border}(a), we note that there is also an evidence of a
sharp transition from the nonconservative to the conservative regimes
of the model, for which the border dependence of epicenters is
radically different. 

The crucial role of the border in this model was already pointed out
by Middleton and Tang~\cite{middleton:1995}, who argued that the inhomogeneity
introduced by the open boundary inhibits the synchronization of the
bulk, which would otherwise reach a periodic state, as it happens with
the system with periodic boundaries. The resulting
``self-organization'' would begin at the border and then proceed
towards the bulk, following a power-law in time. Also, it has been
shown in~\cite{lise:2001}, that while the statics of event sizes in the OFC
model does not seem to obey Finite Size Scaling (FSS), this behavior
is recovered only when events inside a smaller internal subset of the
tension lattice are considered. Thus, the existence of non-scaling
border effects is to be expected. 

We proceed to examine the dynamics of the epicenters in the model, but
only those unrelated to the non-scaling border effect. Therefore,
unless otherwise noted, we ignored all the epicenters belonging to an
outer layer of $100$ sites in the lattice, for all the systems
studied. 

\section{Sequences of epicenters as a complex network} 
\label{sec:epicenter-graph}

A graph (or network) is a set of discrete items, called vertices or nodes, with
connections between them, called edges or links. An edge, connecting
vertices $i$ and $j$, is \emph{directed} if it is defined in only one
direction (connects vertex $i$ with vertex $j$, for instance, but not
site $j$ with site $i$) and a graph is said to be directed if its
edges are directed. There may be more than one edge between a pair of
vertices, and the graph is called in this case a \emph{multigraph}. 
The number of edges connected to a vertex is called the \emph{degree}
of the vertex; since there may be more than one edge between two
vertices, the degree of a vertex is not necessarily equal to the
number of its neighbors. If the graph is directed, it is then possible
to talk about \emph{out-degree} (number of edges leaving a vertex) and
\emph{in-degree} (number of edges incident to a vertex). The degree
distribution of a graph, $P(k)$, gives the probability that a randomly
sampled vertex has degree $k$. Graphs have been the subject of
systematic study by mathematicians for some time, but recent years
witnessed a growth in the interest on this subject among physicists,
with emphasis on large-scale statistical properties of graphs. Many
statistical mechanics concepts and techniques have been widely used,
and a good review on recent developments in this subject can be found
in~\cite{newman:2002}.  We will show that some tools of network theory
can help to get a deeper understanding of the dynamics of the OFC
model and maybe of the dynamics of real earthquakes.

\begin{figure}
  \centering
    \psfrag{1}[][][2.8]{1}
    \psfrag{2}[][][2.8]{2}
    \psfrag{3}[][][2.8]{3}
    \psfrag{4}[][][2.8]{4}
    \psfrag{5}[][][2.8]{5}
    \psfrag{6}[][][2.8]{6}
    \psfrag{7}[][][2.8]{7}
    \psfrag{8}[][][2.8]{8}
    \psfrag{9}[][][2.8]{9}
    \psfrag{10}[][][2.8]{10}
    \psfrag{11}[][][2.8]{11}
    \psfrag{12}[][][2.8]{12}
    \psfrag{13}[][][2.8]{13}
    \psfrag{14}[][][2.8]{14}
    \psfrag{15}[][][2.8]{15}
    \psfrag{16}[][][2.8]{16}
    \psfrag{17}[][][2.8]{17}
    \psfrag{18}[][][2.8]{18}
    \psfrag{19}[][][2.8]{19}
    \psfrag{20}[][][2.8]{20}
    \psfrag{21}[][][2.8]{21}
    \psfrag{22}[][][2.8]{22}
    \psfrag{23}[][][2.8]{23}
    \psfrag{24}[][][2.8]{24}
    \psfrag{25}[][][2.8]{25}
    \resizebox{0.7\columnwidth}{!}{\includegraphics{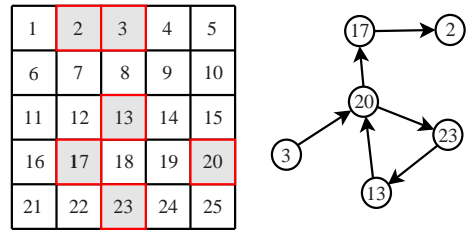}}
    \caption{ (color online) An example of an epicenter graph (right), generated from a 
      sequence of epicenters (marked in red) from a 5x5 tension lattice
      (left). The graph corresponds to the following sequence of
      epicenters: 3, 20, 23, 13, 20, 17, 2. \label{fig:dirmultigraph}}  
\end{figure}

The sequence of epicenters in the OFC model can be used to construct a
directed multigraph in the following manner. Each site that is an
epicenter represents a vertex. Two consecutive epicenters are
connected by a directed edge, from the first to occur to the second
(see Fig.~\ref{fig:dirmultigraph}). Since, in principle, the same
site can become an epicenter two times consecutively, loops are
allowed (but don't occur often). It is also possible for the same
sequence of epicenters to happen more than once, so parallel edges are
also allowed. This graph has certain regularities: The out-degree of
every vertex is always equal to the in-degree, except for the very
first and last epicenters of the sequence, and therefore the total
degree is always an even number. Also, if the direction of the edges
is ignored, the graph is always composed of only one component. 

We have constructed graphs for the epicenters of the OFC model with
$L=400$ and $800$, and for $\alpha=0.25$, $\alpha=0.22$ and
$\alpha=0.18$. We also considered the graphs for epicenters of
different earthquakes sizes. We then observed the degree distribution
and the degree correlation of the graph. The results for the
nonconservative regime are averages over 5 to 11 graphs, depending on
the size of earthquakes considered, each graph with $6\times10^6$
edges. 

\subsection{Degree distribution}
Since the in-degree of the network is equal to the out-degree, it is
sufficient to describe only one of the two, and here we choose
arbitrarily the in-degree. 

\begin{figure}
  \centering
  \psfrag{b0}[ct][rt][1.2]{$b=0$}
  \psfrag{b100}[c][r][1.2]{$b=100$}
  \psfrag{0}[l][l][1.5]{$s \ge 1$}
  \psfrag{2}[l][l][1.5]{$s \ge 2$}
  \psfrag{5}[l][l][1.5]{$s \ge 5$}
  \psfrag{30}[l][l][1.5]{$s \ge 30$}
  \psfrag{0.25}[c][c][1.5]{(a) $L=400$, $\alpha=0.25$}
  \psfrag{0.18}[c][c][1.5]{(b) $L=800$, $\alpha=0.18$}
  \psfrag{0.22}[c][c][1.5]{(c) $L=800$, $\alpha=0.22$}
  \psfrag{2.2(1)}[r][r][1.2]{$\gamma=2.2(1)$}
  \psfrag{2.8(1)}[r][r][1.2]{$\gamma=2.8(1)$}
  \resizebox{1.0\columnwidth}{!}{\includegraphics{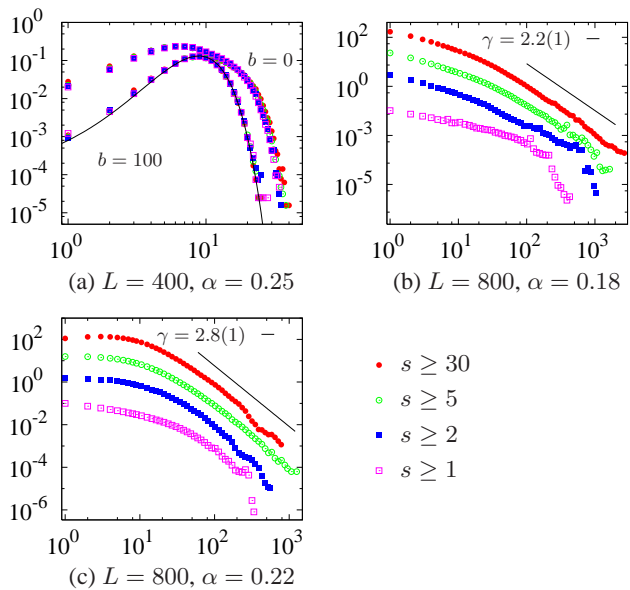}}
  \caption
  {(color online) In-degree distribution $P(k)$ (y axis) in function
    of the in-degree $k$ (x axis), for different values of $L$,
    $\alpha$, and earthquake sizes $s$. In (a) are shown the
    distributions for two different sizes of the discarded border $b$,
    and the solid line is the corresponding Poisson distribution. The
    data for $b=0$ was shifted upwards for clarity. In (b) and (c) the
    solid line is the result of fitting a power-law $k^{-\gamma}$ to
    the data when $s\ge 30$. The data for different earthquake sizes
    were shifted upwards for clarity. All quantities are dimensionless.
    \label{fig:deg-dist}}
\end{figure}

For the conservative regime (Fig.~\ref{fig:deg-dist}(a)), the
in-degree distribution seems to be a Poisson (which gets stretched if
more sites from the border are considered), indicating that, in this
regime, epicenters in the bulk of the lattice occur randomly. 
Moreover, the degree distribution does not depend on the minimum size
of the earthquakes considered. 

For the nonconservative regime the situation changes. As can be seen
in Fig.~\ref{fig:deg-dist}(b) and~\ref{fig:deg-dist}(c), if only
larger earthquakes are considered, the in-degree distribution
resembles more a power-law. The exponent of the power-law seem to be
dependent on $\alpha$, with smaller $\alpha$ leading to steeper lines. 
For $\alpha=0.22$ and $s\ge30$, in Fig.~\ref{fig:deg-dist}(c), the
high fluctuations at the tail of the in-degree distribution represent
a lack of statistics, due to an average over only five realizations of
the graph, while for $s\ge5$, for instance, the average was over ten
different graphs.  For both $\alpha=0.22$ and $\alpha=0.18$, the
difference of inclination of the power-law region of the distributions
is very small between the data for $s\ge5$ and $s\ge30$, indicating
that it is not strongly dependent on the lower bound of the considered
earthquake sizes, provided it is large enough for the power-law to
emerge. 

\subsection{Correlations between degree distribution and tension
  distribution in the lattice}

\begin{figure}
  \centering
  \resizebox{0.7\columnwidth}{!}{\includegraphics{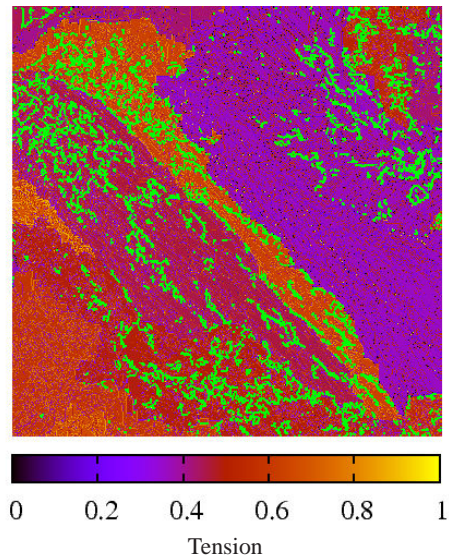}}\\
  Tension
  \caption{ (color online) Snapshot of the tension lattice at the
    stationary state, for $L=800$ and $\alpha=0.18$. The next $10^4$
    epicenters, for earthquake sizes $s\ge2$, after this
    configuration, are marked in green. All quantities are
    dimensionless. \label{fig:800-0.18-tension-corr-2-inf-b100}}
\end{figure}

It is interesting to observe where the epicenters happen in the
tension lattice. As has already been shown in~\cite{drossel:2002}, the
stationary state of the OFC model, for $\alpha<0.25$
(nonconservative), exhibits patchy synchronized regions within the
bulk of the system with sites that have similar tension, and behave
similarly to the OFC model with periodic boundary conditions,
exhibiting heavy synchronization. As can be seen in
Fig.~\ref{fig:800-0.18-tension-corr-2-inf-b100}, for $\alpha=0.18$,
the epicenters seem to happen mostly in the frontiers among those
synchronized regions, and in valley-like structures inside the
plateaus. As only larger earthquakes are considered, the epicenters
happen increasingly in smaller and less structured regions (not
shown). The same behavior was also observed for $\alpha=0.22$.

\begin{figure}
  \centering
  \resizebox{0.7\columnwidth}{!}{\includegraphics{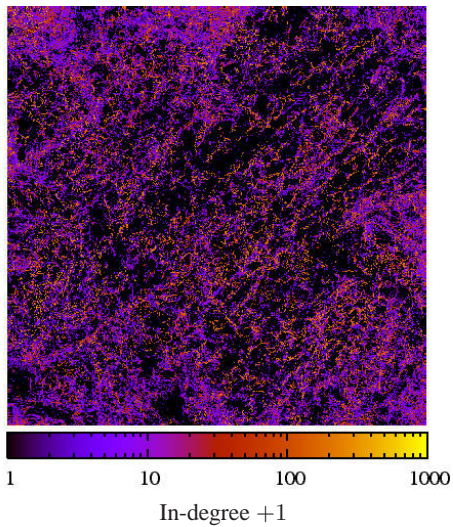}}\\
  In-degree $+1$
  \caption{ (color online) In-degree of vertices placed in the bulk of tension lattice, i.e.,
    the number of times a site was an epicenter, for $L=800$ and
    $\alpha=0.18$. Only earthquakes with sizes $s\ge2$ were
    considered. All quantities are dimensionless.
    \label{fig:800-0.18-in-degree-vs-coord-2-inf-b100}}
\end{figure}

In Fig.~\ref{fig:800-0.18-in-degree-vs-coord-2-inf-b100} can be seen the
in-degree of a vertex placed in the tension lattice, i.e., the number
of times a site was an epicenter, for $\alpha=0.18$ and $s\ge2$. The
epicenters seem to be well distributed inside the bulk, but aggregated in
stripe-like structures. For $s\ge30$ the epicenters seem considerably less
aggregated (not shown). For $\alpha=0.22$ the results were observed to
be very similar. 

\subsection{Degree correlation}

One further basic aspect of the epicenter network which we analyzed was
the degree correlation, i.e., how vertices are connected to each other
based on their degrees. We look at the average in-degree of the
nearest ``out-neighbors'' of a vertex (vertices that receive an edge
coming from it), in function of the degree of the vertex. 

\begin{figure}
  \centering
  \psfrag{s0}[l][l][1.5]{$s \ge 1$}
  \psfrag{s2}[l][l][1.5]{$s \ge 2$}
  \psfrag{s5}[l][l][1.5]{$s \ge 5$}
  \psfrag{s30}[l][l][1.5]{$s \ge 30$}
  \psfrag{0.25}[c][c][1.5]{(a) $\alpha=0.25$}
  \psfrag{0.18}[c][c][1.5]{(b) $\alpha=0.18$}
  \psfrag{0.22}[c][c][1.5]{(c) $\alpha=0.22$}
  \resizebox{1.0\columnwidth}{!}{\includegraphics{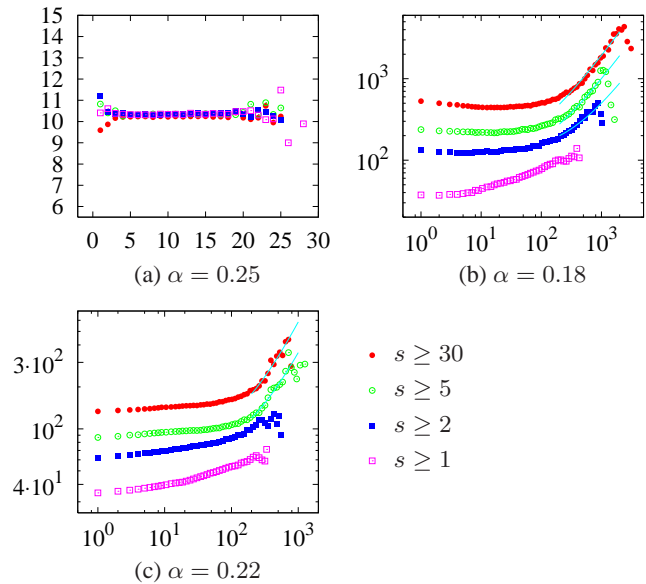}}
  \caption{ (color online) Average in-degree of nearest out-neighbors
    $k_{nn}(k)$ (y axis) in function of the in-degree $k$ (x axis),
    for different values of $L$, $\alpha$, and earthquake sizes
    $s$. The solid lines are fitted straight lines. All quantities are
    dimensionless.\label{fig:deg-corr}}
\end{figure}

We found that for the conservative regime
(Fig.~\ref{fig:deg-corr}(a)), the graph seems to be uncorrelated,
with the in-degree of the nearest neighbors being independent on the
in-degree of the originating vertex. Together with the in-degree
distribution (a Poisson), this puts this graph closer to the class of totally
random graphs such as the Erd\H{o}s-R\'{e}nyi graph~\cite{erdos:1959}. 

The situation is again very different for the nonconservative regime
(Figs.~\ref{fig:deg-corr}(b) and~\ref{fig:deg-corr}(c)). In that
case, the degrees seem highly correlated, with vertices with high
in-degree connecting predominantly to other vertices of high
in-degree, which makes the network assortative. The correlation seems
to be linear for higher degrees, when only larger earthquakes are
considered~\footnote{The in-degree correlation of the graph for
  $s\ge1$ is also crescent, and perhaps also linear. But the lack of
  vertices of high degree makes it difficult to be certain.}. Citation
networks~\cite{price:1976,barabasi:1999} and other networks that are
grown with a preferential attachment rule have a quite different
behavior, with an in-degree distribution following a power-law,
but in those cases the degree correlation also decays with a power
law~\cite{pastor-satorras:2001}, converging to a constant value for
large in-degrees. Thus, the dynamics responsible for generating this
network must be fundamentally different than the dynamics generated by
a preferential attachment rule. Recently it has also been found that
a very similar network, when constructed with real earthquake data, is also
assortative and exhibits similar degree correlations~\cite{abe:2006}. 

\begin{figure}
  \centering
  \psfrag{event size}[c][c][1.4]{Event size}
  \psfrag{avg deg}[c][c][1.4]{$\left<k\right>$}
  \psfrag{events}[c][c][1.4]{Events}
  \psfrag{earthquakes}[l][l][1.2]{Events}
  \psfrag{avg-deg}[l][l][1.2]{$\left<k\right>$}
  \resizebox{1.0\columnwidth}{!}{\includegraphics{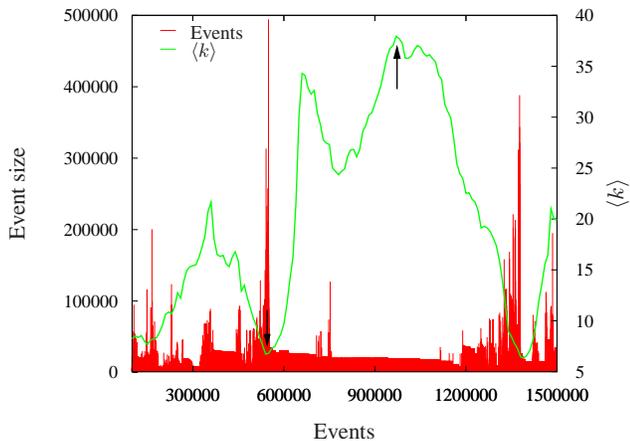}}
  \caption{ (color online) Average in-degree of the subgraph composed
    only of the last $10^5$ events, and the amplitudes of the events
    that generated the graph. The regions indicated by the arrows
    correspond to the subgraphs in Figs.~\ref{fig:graph-attractor-2}
    and~\ref{fig:graph-attractor-1}. All quantities are
    dimensionless.\label{fig:attractor-0.18}}
\end{figure}

What indeed is unveiled by this high correlation amongst high
in-degree vertices is an attracting dynamics: Connections from
vertices of one type are much more probable to vertices of the same
type, eventually trapping the sequence of epicenters in a smaller
region of the lattice, stretching the in-degree distribution, and
generating the observed in-degree correlation.  This trapping seems to
be strongly correlated to the occurrence of very large earthquakes,
and the large scale redistribution of tensions that is caused by them.
This can be seen in Fig.~\ref{fig:attractor-0.18}, where is shown the
average in-degree of the subgraph composed only of the last $10^5$
events, together with the amplitude of the corresponding events.
Whenever a large earthquake occurs, the average in-degree drops,
meaning that the last epicenters happened in a larger number of sites.
In fact the decay of the average in-degree starts before the main big
earthquake, and seems to occur together with the smaller events that
lead up to it, the so called
\emph{foreshocks}~\cite{omori:1894,hergarten:2002,helmstetter:2004}.
Thus, the large events, together with their foreshocks, are
responsible for breaking the attractor, and spreading the epicenters
to a larger region. After the sequence of large events, the trapping
of epicenters starts again, until the next sequence of large events
sweeps it again. Although we did not make an extensive analysis to
define the degree of certitude of the this observation, monitoring the
in-degree of this network may represent a promising way of predicting
an increase in the probability of observing large earthquakes in a
given fault, and to identify, among the small events, the signature of
the foreshocks that preceed a main shock~\footnote{It is important to
  note that the data in Fig.~\ref{fig:attractor-0.18} shows only
  earthquakes that did not initiate inside the discarded outer layer,
  whose magnitude tend to be large, and thus are potentially related
  to the anticipated decay of the average in-degree before a main
  large event.}. Since the network of epicenters generated by the OFC
model seems to reproduce many aspects of the network of epicenters
built from real data~\cite{abe:2004,abe:2006}, including the degree
correlation mentioned above, it would be interesting to see in more
detail if both graphs are actually generated by the same overall
dynamics. This however would require a more systematic and thorough
analysis of real earthquake data, and therefore would be better suited
for a separate work.

\begin{figure}
  \centering
  \resizebox{1.0\columnwidth}{!}{\includegraphics{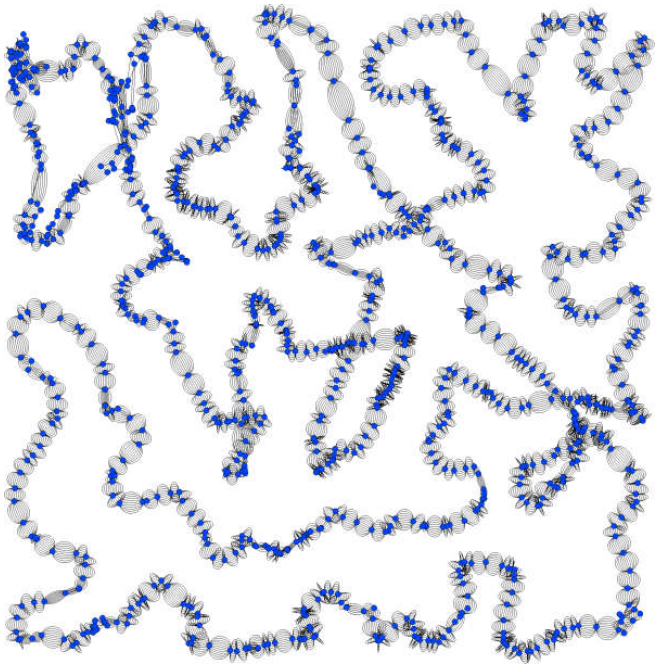}}
  \caption{ (color online) Subgraph composed of $10^4$ consecutive epicenters,
    corresponding to the marked region at the right in
    Fig.~\ref{fig:attractor-0.18}.\label{fig:graph-attractor-2}}  
\end{figure}

\begin{figure}
  \centering
  \resizebox{1.0\columnwidth}{!}{\includegraphics{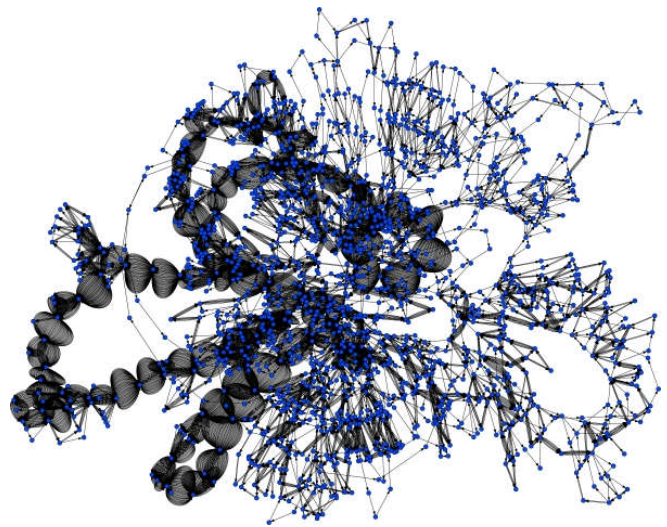}}
  \caption{ (color online) Subgraph composed of $10^4$ consecutive epicenters,
    corresponding to the marked region at the left in
    Fig.~\ref{fig:attractor-0.18}.\label{fig:graph-attractor-1} }  
\end{figure}

To illustrate the topology of the graph during both situations, we show a
subgraph of the whole network, corresponding to a region of $10^4$
events collected during the period  that the dynamics is trapped in an
attractor (Fig.~\ref{fig:graph-attractor-2}), and just after a large
earthquake (Fig.~\ref{fig:graph-attractor-1}), as indicated in Fig.~\ref{fig:attractor-0.18}. As can be
seen in Fig.~\ref{fig:graph-attractor-2}, the attractor region is
dominated by synchronization, where the same sequence of $\sim 10^3$
epicenters occur repeatedly. During the occurrence of the large
events, the same subgraph looks like
Fig.~\ref{fig:graph-attractor-1}, where synchronization is still
present, but in a much smaller degree. 

\section{Markov networks with hidden variables}
\label{sec:markov-graph}

In this section we describe a general random graph model, based on
hidden variables and a Markov Chain. It is based on a similar class of
networks developed by Bogu\~{n}\'{a} et al~\cite{boguna:2003}, but
modified in order to account for the topology of the epicenter graph
observed in the OFC model.  Our goal is to better understand the type
of dynamics that is able to generate graphs with properties examined
in the previous section. 

Consider a set of N vertices, where $N \gg 1$. To each vertex $\nu$ is assigned a
hidden variable $h_\nu$, sampled from a distribution $\rho(h)$. A
directed multigraph can be constructed by a Markov chain, in the
following manner: Starting from a random vertex $\mu$, a directed edge
is added from $\mu$ to $\nu$ with probability $P(\mu \rightarrow \nu)
\equiv r(h_\mu,h_\nu)$, and likewise from $\nu$ to any other vertex
$\omega$ with probability given by $r(h_\nu,h_\omega)$, and so forth. 
After a transient stage, the graph will have properties that are
entirely defined by $\rho(h)$ and $r(h,h')$. This graph is rather
general, and, in fact, every Markov process generates such a graph if
the discrete states of the chain are thought as vertices and the
transition as directed edges. With this basic procedure in mind we can
proceed to calculate the statistical properties of the graph. 

\subsection{In-degree distribution}

Like in the  network of epicenters, every vertex of the network
generated in the way described above has the
in-degree equal to the out-degree. Thus, it is sufficient to describe
only one of the two.  To find the in-degree distribution of this
graph, one must consider an ensemble of graphs and the probability in
the ensemble of one vertex $\nu$ receiving one connections after a
time $T$, $w_\nu(T)$, which is given by
\begin{equation}
  \label{eq:wnu}
  w_\nu(T+1) = \sum_\mu P(\mu \rightarrow \nu) w_\mu(T). 
\end{equation}
After a long time $T$, the system reaches the stationary state,
$w(\infty)$, given by
\begin{equation}
  w(\infty) = \mathbf{P}^n w(\infty),
\end{equation}
where $\mathbf{P}$ is the transition matrix defined by $P(\mu
\rightarrow \nu)$,  $w(T)$ is the state vector at time $T$, and $n$ is
the period of the solution (we will consider only $n=1$ from now on). 

The probability that a vertex $\mu$ has in-degree $k$ after a time
$T\gg 1$, $P(k|\mu,T)$, is given simply by the binomial distribution,
\begin{equation}
  \label{eq:pkmu}
  P(k|\mu,T)=\binom{T}{k}w_\mu^k(1-w_\mu)^{T-k} \approx \frac{(Tw_\mu)^k e^{-Tw_\mu}}{k!},
\end{equation}
where $w_k \equiv w_k(\infty)$, which can be approximated by the
Poisson distribution, as in the rightmost term. 

The total in-degree distribution after a time $T$, $P(k|T)$, is then given by
\begin{equation}
  \label{eq:pkt}
  P(k|T)= \frac{1}{N}\sum_\mu P(k|\mu,T). 
\end{equation}

Now since a vertex $\mu$ is labeled uniquely by its hidden variable
$h_\mu$, we must have then that $w_\mu\equiv w(h_\mu)$. Thus, $w(h)$
can be obtained by rewriting equation~\ref{eq:wnu},
\begin{equation}
  w(h)=N\int \limits_h r(h_\mu,h)w(h_\mu)\rho(h_\mu)dh_\mu,
\end{equation}
assuming that $h$ is a continuous variable (the last expression would
just be a sum if it were discrete). Solving this integral
equation for $w(h)$, it is possible then to obtain the degree distribution through
equation~\ref{eq:pkt},
\begin{equation}
  \label{eq:pk}
  P(k,T) = \int \limits_h  \frac{\left(Tw(h)\right)^k e^{-Tw(h)}}{k!}\rho(h)dh. 
\end{equation}

\subsubsection{In-degree correlation}

It is also possible to calculate the degree correlation of this graph. 
The probability of one vertex $\mu$, with in-degree $k$, connecting to
another vertex of degree $k'$ is given by
\begin{equation}
  \label{eq:pkk'mu}
  P(k'|k,\mu,T) = \frac{P(k|\mu,T)}{NP(k,T)}\sum_\nu P(\mu\rightarrow\nu)P(k'-1|\nu,T). 
\end{equation}
The total probability of one vertex with degree $k$ connecting to one
of degree $k'$ is then simply
\begin{equation}
  \label{eq:pkk'}
  P(k'|k,T) = \sum_\mu P(k'|k,\mu,T),
\end{equation}
and the average in-degree of the nearest out-neighbors is just then 
\begin{equation}
  \label{eq:knn}
  \bar{k}_{\text{nn}}(k,T)=\sum_{k'}k'P(k'|k,T). 
\end{equation}
In terms of the hidden variables, substituting equations~\ref{eq:pk}
and~\ref{eq:pkmu} in~\ref{eq:pkk'} and calculating the sum in~\ref{eq:knn}, we have then,
\begin{multline}
  \label{eq:knnh}
  \bar{k}_{\text{nn}}(k,T) = 1 +
  \frac{N}{P(k,T)}\iint \limits_{h}
  \frac{\left(Tw(h_\mu)\right)^ke^{-Tw(h_\mu)}}{k!} \\
  \times r(h_\mu,h_\nu)Tw(h_\nu)\rho(h_\mu)\rho(h_\nu)dh_\mu dh_\nu. 
\end{multline}

\subsubsection{Attractor dynamics}

\begin{figure}
  \captionsetup[subfloat]{captionskip=7pt}
  \centering
   \psfrag{hn}[c][c][2.0]{$h_\nu$}
   \psfrag{hm}[tc][tc][2.0]{$h_\mu$}
   \psfrag{r}[l][c][2.0]{$r(h_\mu,h_\nu)$}
   \psfrag{dec}[c][c][2.0]{$\frac{G(h_\nu)}{\rho(h_\mu)}$}
   \psfrag{pref}[c][c][2.0]{$h_\nu^\gamma$}
   \subfloat[]{\resizebox{0.4\columnwidth}{!}{\includegraphics{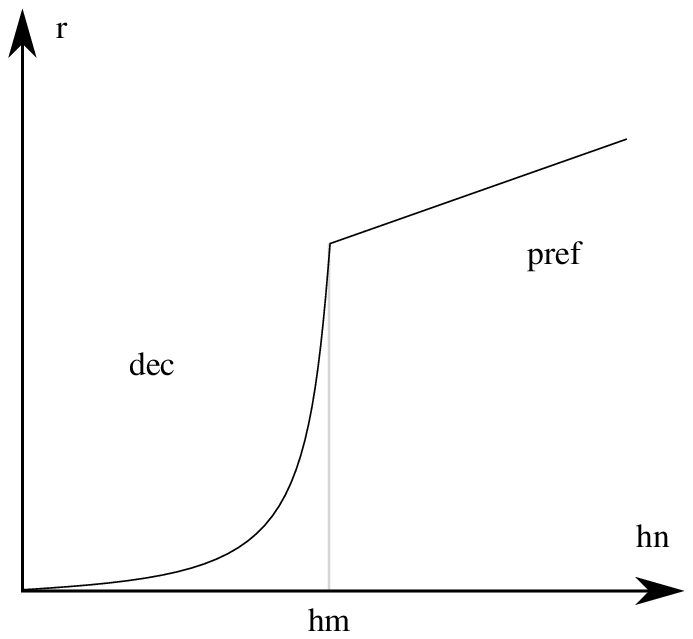}}}
   \qquad
   \psfrag{r}[l][c][2.0]{$r(h_\mu,h_\nu)\times \rho(h_\nu)$}
   \psfrag{dec}[c][c][2.0]{$G(h_\nu)$}
   \psfrag{pref}[c][c][2.0]{$h_\nu^\gamma\rho(h_\nu)$}
   \subfloat[]{\resizebox{0.4\columnwidth}{!}{\includegraphics{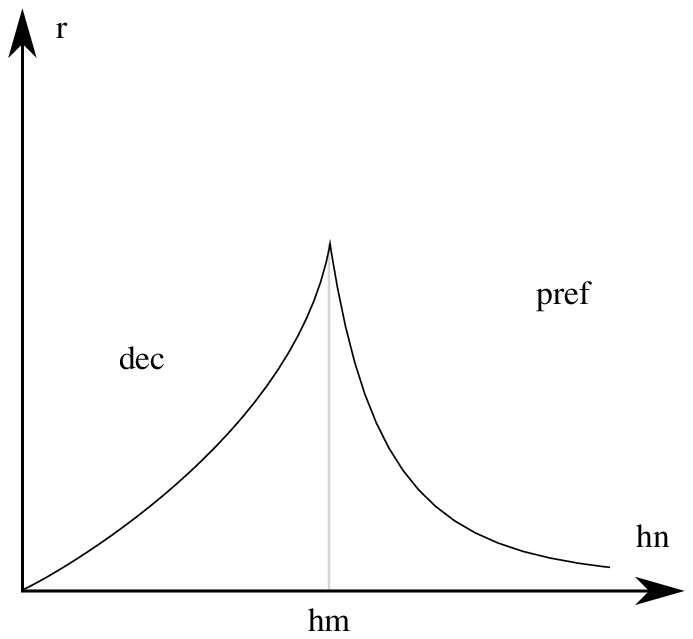}}}
  \caption{(a) Connection probability (equation~\ref{eq:rhh}) from a vertex
    $\mu$ to a vertex $\nu$, and (b) connection probability from a
    vertex $\mu$ to any vertex with hidden variable $h_\nu$. \label{fig:rhh}}
\end{figure}

We want to understand how correlations such as seen in
Figs.~\ref{fig:deg-corr}(b) and~\ref{fig:deg-corr}(c), and
power-law distributions can arise from this type of network. For that
we must define a suitable $r(h,h')$ and $\rho(h)$. It is clear that
what uniquely defines the in-degree of some vertex is its hidden
variable. Thus, for the in-degree correlation to be of the form
$\bar{k}_{\text{nn}}(k) \sim k$ for large $k$, we must have that
$\bar{h}'(h)\sim h$ for large $h$, where $\bar{h}'(h)$ is the average
hidden variable of the out-neighbors of a vertex with hidden variable
$h$.  With this in mind, we define then the following general
expression for the connection probability,
\begin{multline}
  \label{eq:rhh}
  r(h_\mu,h_\nu) =
  F(h_\mu)\left[\frac{G(h_\nu)h_\mu^\gamma}{\rho(h_\nu)}[h_\nu<h_\mu]\right. \\
    \left. + \frac{G(h_\mu)h_\nu^\gamma}{\rho(h_\mu)}[h_\nu>h_\mu] \right]
\end{multline}
Where $G(h)$ is a function that dictates how fast the connection probability
decays for $h_\nu < h_\mu$ (see Fig.~\ref{fig:rhh}), and the
exponent $\gamma$ defines the preference with which vertices with
higher $h$ are chosen. The function $F(h_\mu)$ is simply given by
the normalization condition $\sum_\nu r(h_\mu,h_\nu) = 1$. 

We considered a few shapes for $G(h)$ and $\rho(h)$ and calculated the
degree distribution and degree correlation through
equations~\ref{eq:pk} and~\ref{eq:knnh}, always  for $k\gg1$. The results
are summarized in table~\ref{tab:pk_knn}. 

\begin{table}
  \begin{tabular}{|c|c|c|c|}\hline
    $\rho(h)$                    & $G(h)$      & $\bar{k}_\text{nn}(k)$ & $P(k)$ \\ \hline\hline
    $\displaystyle\frac{h^{-\beta}}{\beta-1}$ & $1$ & $\sim k$ & $\sim k^{-\frac{\beta+\gamma}{\gamma+1}}$ \\ \hline
    $\beta e^{-\beta h}$  & $1$  & $\sim k$ & $\sim k^{-\frac{\gamma}{\gamma+1}}e^{-C k^\frac{1}{\gamma+1}}$ \\\hline
    $\beta e^{-\beta h}$  & $e^{\xi h}$ &  $\sim k$ & $\sim  k^{-\frac{\beta+\xi}{\xi}}$ \\\hline
  \end{tabular}
  \caption{Different asymptotic shapes for $\bar{k}_\text{nn}(k)$ and
    $P(k)$ for different shapes of $G(h)$ and $\rho(h)$, for $k\gg
    1$.\label{tab:pk_knn}} 
\end{table}

What we find is that the effect of adopting a connection probability
like the one described by
equation~\ref{eq:rhh} is to generate a in-degree distribution
corresponding to a stretched form of $\rho(h)$. If $G(h)$ is
independent of $h$, and the trapping in the region of similar $h$ is
the weakest, we have the following possibilities: If $\rho(h)$ is a power-law with
exponent $\beta$, then $P(k)$ will also be a power-law with exponent
in the region $[1,\beta]$, approaching $1$ if $\gamma$ is large. When
$\rho(h)$ is a exponential distribution, the resulting in-degree
distribution will be a stretched exponential as indicated in
table~\ref{tab:pk_knn}, which will also resemble a power-law if $\gamma$ is
relatively large. Now, considering a stronger ``trapping effect''  with
$G(h)$ increasing exponentially, we have that an exponential
$\rho(h)$, with decay parameter $\beta$, is enough to create a power-law
distribution of in-degrees, with exponents in the interval
$[1,\beta]$, approaching $1$ with faster $G(h)$. This means that it is
not necessary to assume an intrinsic scale-invariance, represented by
a power-law in $\rho(h)$, for the existence of a power-law in $P(k)$. 
Furthermore, the asymptotic in-degree distribution in this case does
not depend on $\gamma$, being totally dominated by the ``trapping''
behavior, and not by the preference of connection. 


The process described above shows a variety of ways in which graphs
with in-degree distributions resembling power-laws and linear
in-degree correlation can be created. Looking at only these
properties, it is not possible to know which one of the possibilities
(if any) is more likely to describe the epicenter network. Moreover,
the process above would not account for the strong synchronization
observed in Figs.~\ref{fig:graph-attractor-2}
and~\ref{fig:graph-attractor-1}. After all,  the sequence of
epicenters are probably not simple Markovian processes.  However, the
above model, as a first approximation,
serves the purpose of illustrating  how such correlations and
in-degree distribution can come to place, and presents a general
analytical framework for further modeling. 

\section{Conclusions}
\label{sec:conclusions}

We have shown that the epicenters in the OFC model occur predominantly
near the boundary of the lattice, but this preference does not seem to
scale with system size. This border affinity depends on the
dissipation parameter $\alpha$, being thinner for smaller values of
$\alpha$. It is also dependent on the earthquake size, with epicenters
of larger earthquakes having a border effect which decays more slowly
towards the bulk.  We have also studied the network of consecutive
epicenters, and found that it is sharply different in the two regimes
of the model: In the conservative regime it is rather featureless,
with uncorrelated in-degree statistics and Poisson in-degree
distribution. However, in the nonconservative regime, it has an
unusual linear degree correlation amongst vertices of high degree, and
a broad distribution of in-degrees resembling a power-law, but only
when the smaller earthquakes are not considered. The in-degree
distribution and correlation in this regime is similar to what was
found very recently for real earthquakes~\cite{abe:2004,abe:2006}.
Furthermore, we noticed that the high correlation of in-degrees is due
to an attractor dynamics where the occurrence of epicenters tend to
synchronize, with the same sequence of epicenters occurring
continuously. This synchronization is broken by large earthquakes,
which spread the epicenters over a larger portion of the lattice, thus
populating the graph with vertices of smaller in-degree.
Interestingly, the effects of the large events on the topology of the
epicenter network are noticeable before the actual main event, and
seem to be related to the series of increasingly larger foreshocks
that precede it. Since the prediction of the OFC model that there
would be an in-degree correlation in the epicenter graph corresponds
to what has been recently found for real earthquakes~\cite{abe:2006},
further detailed analysis of this behavior may prove useful for the
prediction of large earthquakes. Lastly we described a general
analytical network model based on a Markovian process and hidden
variables, which is able to reproduce the most general aspects of the
epicenter network, when a suitable attractor dynamics is specified.
There are several aspects of the dynamics of epicenters that remain
uncovered.  It would be of special interest to look at other
topological properties of the epicenter graph, such as the dependence
of the clustering coefficient on in-degree, and the existence of
community structure~\cite{newman:2004,muff:2005}.  Furthermore it
would also be useful to compare in detail some of the results here
obtained, such as the dynamics responsible for the in-degree
correlation and the epicenter synchronization, with the epicenter
network of real earthquakes.

\begin{acknowledgments}
This work was supported by Funda\c{c}\~{a}o de Amparo \`{a} Pesquisa
do Estado de S\~{a}o Paulo (FAPESP), process number 03/03429-6. 
\end{acknowledgments}

\bibliography{bibliography}

\end{document}